\documentclass[twocolumn,prl,aps]{revtex4-1}
\usepackage{graphicx}
\usepackage{epsfig}
\usepackage{amsmath}
\newcommand{\LZL}{ \lambda_\textrm{{zero,lab}} }
\newcommand{\LZ}{ \lambda_\textrm{{zero}} }
\begin{document}
\title{Atom Interferometer Gyroscope with Spin-Dependent Phase Shifts Induced by Light near a Tune-out Wavelength}
\author{Raisa Trubko$^{1}$, James Greenberg$^{2}$, Michael T. St.Germaine$^{2}$, Maxwell D. Gregoire$^{2}$,  William F. Holmgren$^{2}$, Ivan Hromada$^{2}$, and Alexander D. Cronin$^{1,2}$}
\affiliation{$^{1}$College of Optical Sciences University of Arizona, Tucson, Arizona 85721, USA}
\affiliation{$^{2}$Department of Physics, University of Arizona, Tucson, Arizona 85721, USA}
\date{April 1, 2015}
\begin{abstract}
Tune-out wavelengths measured with an atom interferometer are sensitive to laboratory rotation rates because of the Sagnac effect, vector polarizability, and dispersion compensation.  We observed shifts in measured tune-out wavelengths as large as 213 pm with a potassium atom beam interferometer, and we explore how these shifts can be used for an atom interferometer gyroscope.
\end{abstract}
\maketitle

Atom interferometers have an impressive variety of applications ranging from inertial sensing to measurements of fundamental constants, measurements of atomic properties, and studies of topological phases \cite{Cro09}.  In particular, making a better gyroscope has been a long standing goal in the atom optics community because atom interferometers have the potential to outperform optical Sagnac gyroscopes.  Advances in the precision and range of applications for atom interferometry have been realized by using interferometers with multiple atomic species  \cite{Var09, Bon13, Ber07, Hol10}, multiple atomic velocities \cite{Gus00,Dur06, Can06,Dic13, Ham95},  multiple atomic spin states  \cite{Sch94,Pet13,Lom14}, and multiple atomic path configurations \cite{Gei11,Rob10,Li14,Gup02,Aok01}.  Here, we use atoms with multiple spin states to demonstrate a new method for rotation sensing. Our atom interferometer gyroscope reports the absolute rotation rate $\Omega$ in terms of an optical wavelength, using a spin-dependent phase echo induced by light near a tune-out wavelength.

\begin{figure}[b]
\begin{center}
\includegraphics[width=8.5cm]{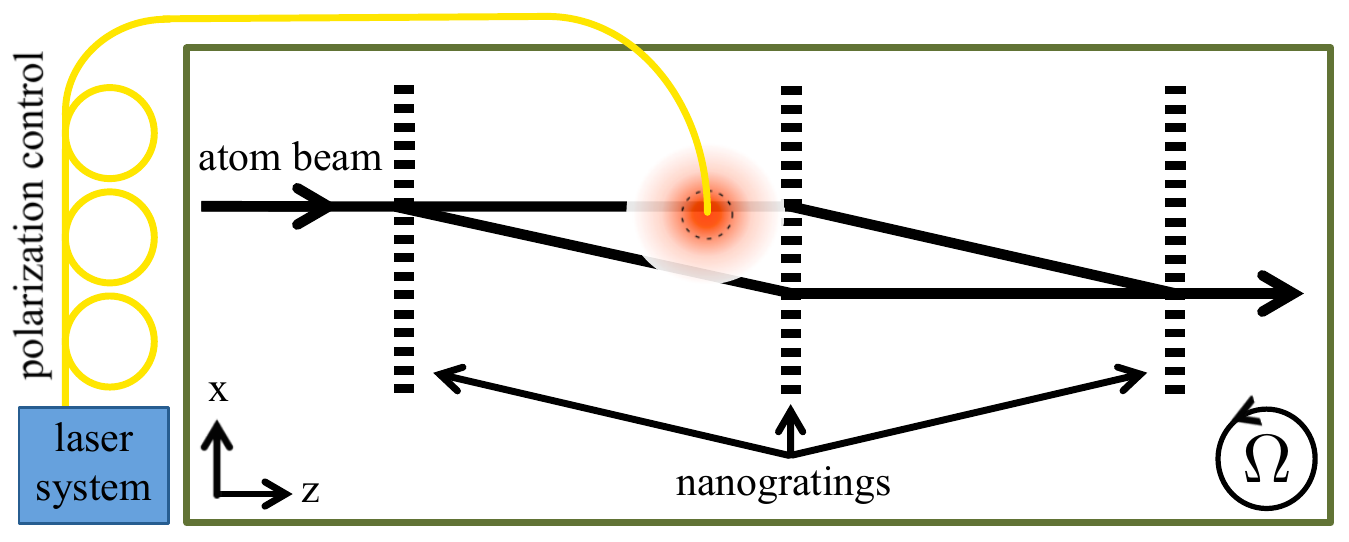}
\caption{(Color online) Apparatus diagram.  The branches of a 3-nanograting Mach-Zehnder atom interferometer \cite{Cro09} are illuminated asymmetrically by laser light propagating perpendicular to the page. An optical cavity (not shown) recycles the light to increase the phase shift. A single mode optical fiber (yellow) guides the laser light into the atom beam vacuum chamber, and the loops in the fiber are used to control the optical polarization.}
\label{fig:app}
\end{center}
\end{figure}

A tune-out wavelength, $\LZ$, occurs where the dynamic polarizability of an atom changes sign between two resonances \cite{Aro11,LeB07,Her12, Che13,Mit13,Top13,Jia13,Jia13_2,Saf13,Saf12_2,Hol12}. Since atomic vector polarizability depends on spin \cite{man86,Ros09,Kie13}, theoretical tune-out wavelengths usually describe atoms with spin $m_F=0$.  The same $\LZ$ should be found, on average, for atoms in a uniform distribution of spin states.  However, in this Letter, we show that the Sagnac effect breaks the symmetry expected from the vector polarizability in a way that makes tune-out wavelengths remarkably sensitive to the laboratory rotation rate.  We measured tune-out wavelengths $\LZL$ using a potassium atom interferometer shown in Fig.~\ref{fig:app} and circularly polarized light, and found that our measurements were shifted by 0.213 nm from the theoretical tune-out wavelength of $\LZ = 768.971$ nm \cite{Aro11}.  This shift is more than 100 times larger than the uncertainty with which $\LZ$ can be measured \cite{Hol12}, and this suggests the possibility of creating a sensitive gyroscope using tune-out wavelengths.  The purpose of this Letter is therefore to explain how an atom interferometer gyroscope can measure the laboratory rotation rate $\Omega$ with the aid of atomic spin-dependent phase shifts induced by light near a tune-out wavelength.  This is a new application of tune-out wavelengths and a new method for atom interferometry that could improve sensors needed for navigation, geophysics, and tests of general relativity.

Atom interferometer gyroscopes \cite{Len97,Gus00,Gus97,Dur06,Can06,Lan12,Dic13,mul09,Wu07,Sto11,Cro09,Bar14} can sense changes in rotation rate ($\Delta \Omega$) because of the Sagnac effect. Some atom interferometers \cite{Gus97,Dur06,Can06,Lan12,Dic13} can also report the absolute rotation rate ($\Omega$) with respect to an inertial frame of reference since the Sagnac phase depends on atomic velocity.  Because the Sagnac phase is dispersive, $\Omega$ can affect the interference fringe contrast.  References \cite{Gus97,Gus00,Dur06,Can06,Lan12,Dic13} applied auxiliary rotations to an atom interferometer to compensate for the earth's rotation $\Omega_{e}$ and thus maximize contrast.  References \cite{Gus97} and \cite{Lan12} even used  contrast as a function of applied rotation rate  in order to measure $\Omega_e$.

In comparison, here we demonstrate optical and static electric field gradients that compensate for dispersion in the Sagnac phase. This is a general example of dispersion compensation \cite{Rob04,Jac08} in which one type of phase compensates for dispersion in another.  Furthermore, we show that circularly polarized light at $\LZ$ makes an observable $\Omega$-dependent phase shift for our unpolarized atom beam interferometer.  This works because spin-dependent dispersion compensation causes higher contrast for one spin state.  Thus, using spin as a degree of freedom and light near a tune-out wavelength, we made a gyroscope that reports the absolute rotation rate $\Omega$ in terms of a light-induced phase shift.

Our gyroscope  uses material nanogratings which permit interferometry with distributions of atomic spin and velocity, both of which are needed in order to cause the shifts in $\LZL$ that are sensitive to $\Omega$.  An atom interferometer like ours was previously shown to monitor changes in rotation rate $\Delta\Omega$ \cite{Len97}. We now show that an atom interferometer gyroscope with material nanogratings can measure absolute rotation rates smaller than $\Omega_e$. This is significant because nanogratings offer some advantages such as simplicity, reliability, and spin-independent and nearly velocity-independent diffraction amplitudes that may enable more robust and economical $\Omega$ sensors.

We studied the light-induced phase shift $\phi$ for an ensemble of atoms, which we model as
\begin{equation}
\phi = \phi_{\textrm{on}} - \phi_{\textrm{off}}
\label{eq:phase}
\end{equation}
where $\phi_{\textrm{on}}$ is the measured phase when the light is on and $\phi_{\textrm{off}}$ is the measured phase when the light is off. For an atom beam with a velocity distribution $P_{v}\left(v\right)$ and a uniform distribution of spin states $P_{s}\left(F, m_{F}\right) = 1/8$, the contrast $C_{\textrm{on}}$ and phase $\phi_{\textrm{on}}$ for the ensemble are described by
\begin{equation}
C_{\textrm{on}}e^{i\phi_{\textrm{on}}} = C_{o}\sum_{F,m_{F}}P_{s}\left(F, m_{F}\right)\int_{0}^{\infty}{P_{v}(v)e^{i \Phi_{\textrm{total}}}dv}
\label{eq:cphase}
\end{equation}
where $ \Phi_{\textrm{total}} = \Phi_{L}+\Phi_{S}+\Phi_{a}+\Phi_{o}$.  Here, $\Phi_{L}$ is the velocity-dependent and spin-dependent phase caused by light, $\Phi_{S}$ is the velocity-dependent Sagnac phase, $\Phi_{a}$ is the velocity-dependent phase induced by an acceleration or gravity, $\Phi_{o}$ is the initial phase, and $C_{o}$ is the initial contrast of the interferometer.  A similar equation can be written for $C_{\textrm{off}}e^{i\phi_{\textrm{off}}}$ with the light off, so that $\Phi_{L}=0$. Our atom beam has a velocity distribution $P_{v}(v)$ adequately described by \mbox{$P_{v}(v) = Av^3\textrm{exp}\left[-\left(v-v_{0}\right)^{2}/\left(2\sigma_{v}^{2}\right)\right]$}, where $A$ is a normalization constant \cite{Hab85}.

The Sagnac phase \cite{Len97,Gus97}
\begin{equation}
\Phi_{S} =\frac{4\pi L^{2}\Omega}{vd_{g}}
\label{eq:sag}
\end{equation}
is a function of atomic velocity $v$ and the rotation rate $\Omega$ along the normal of the interferometer's enclosed area. $L$ is the distance between gratings and $d_{g}$ is the period of the gratings. In our interferometer, $d_{g} = 100$ nm and $L = 0.94$ m, so $\Phi_{S} = 2.7$ radians for a 1600 m/s atom beam in our laboratory at $32^{\circ}$ N latitude due to $\mathbf{\Omega_e}$.

\begin{figure}[t]
\begin{center}
\includegraphics[width=9cm]{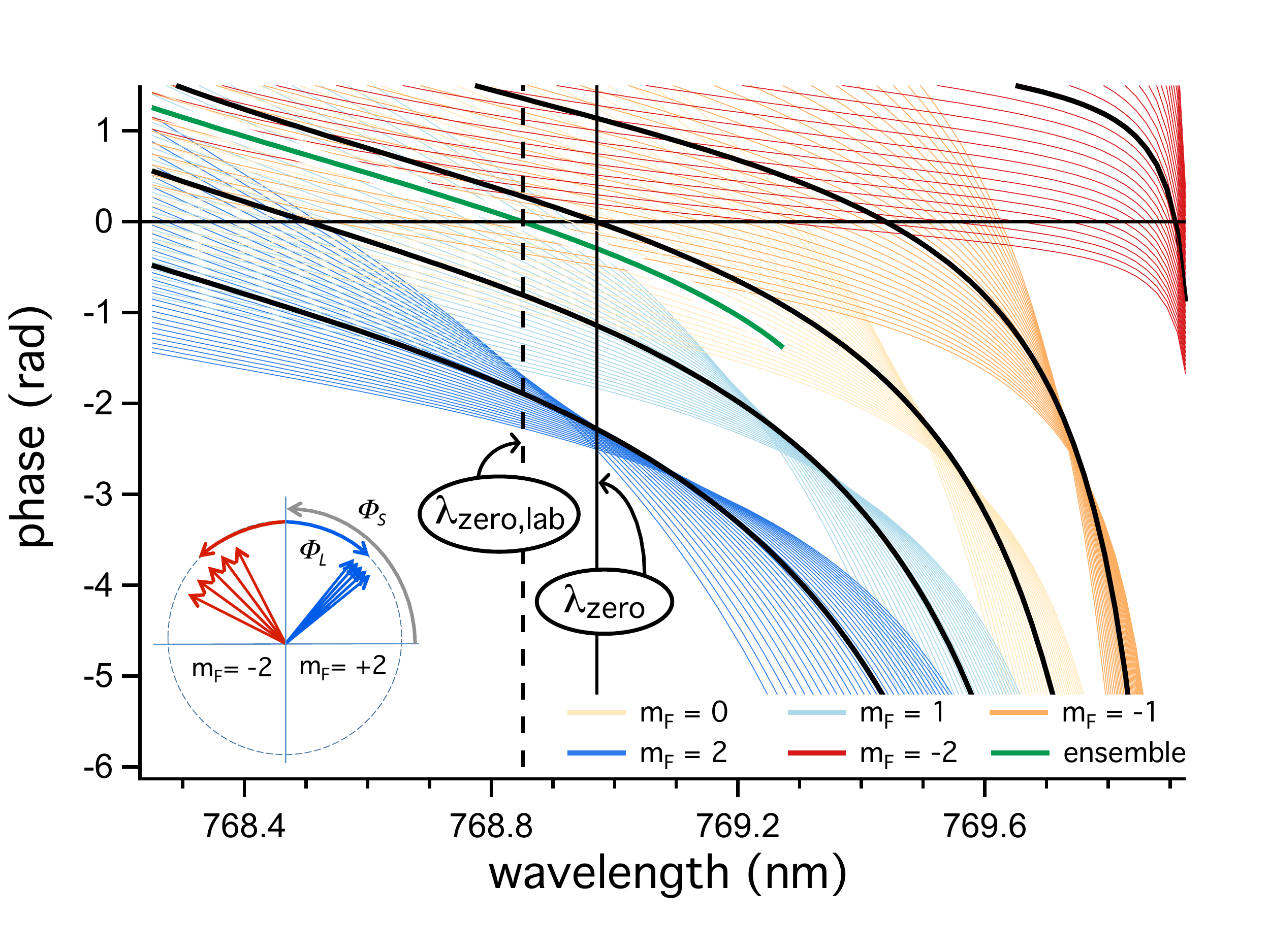}
\caption{Light-induced phase spectra demonstrate dispersion compensation. The phase $\Phi_L(\lambda,v) + \Phi_S(v) - \phi_{\textrm{off}} $  is plotted for 95\% circularly polarized light interacting with 5 atomic spin states (colors) and a range of atomic velocities spanning 80$\%$ to 120$\%$ of $v_0$=2000 m/s.  Black curves show spectra for velocity $v_0$ for each spin state.  Curves for each spin state coalesce in caustics at a different $\lambda$ where spin-dependent $\Phi_L(\lambda,v)$ compensates for dispersion in $\Phi_S(v)$. The ensemble phase shift (green) shows the root in $\phi$ at $\LZL$, which is shifted by $- 120$ pm from $\LZ$. The phasor diagram (inset) illustrates how $\Phi_{L}$ compounds with $\Phi_{S}$ to increase dispersion for one spin state and decrease dispersion for another spin state.}
\label{fig:lzs}
\end{center}
\end{figure}

The gravity phase $\Phi_{a}$ \cite{Len97} is
\begin{equation}
\Phi_{a} = \frac{2\pi L^{2}g\,\text{sin}\left(\theta\right)}{v^{2}d_{g}}
\label{eq:phia}
\end{equation}
where $g\,\text{sin}\left(\theta\right)$ is the gravitational acceleration along the grating wave vector direction. As we discuss later, $\theta$ and $\Phi_{a}$ are small, but non-zero.

The light phase is
\begin{equation}
\Phi_{L} = \frac{\alpha\left(\omega\right)}{2\epsilon_{o}c\hbar v}   \int{s \cdot \left[\frac{d}{dx}I\left(\mathbf{r},\omega\right)\right]  dz}
\label{eq:light}
\end{equation}
where the dynamic polarizability $\alpha(\omega)$ depends on the atomic state $|F,m_F\rangle$ and the laser polarization \cite{man86,Ros09,Kie13}. Near the second nanograting, we shine 50 mW of laser light perpendicular to the plane of the interferometer. The laser's irradiance gradient in a beam with a 100\nobreak\;$\mu$m diameter waist asymmetrically illuminates the atom beam paths as sketched in Fig.~1. The irradiance gradient $\frac{d}{dx} I$ is integrated along the atom beam paths in the $z$-direction.  The path separation $s$ is proportional to $v^{-1}$. Hence, for laser beams much wider than $s$, the light phase $\Phi_L$ approximately depends on $v^{-2}$.  The fact that this does not exactly match the $v^{-1}$ dispersion of the Sagnac phase means the dispersion compensation is imperfect, which is why we see caustics in Fig.~\ref{fig:lzs}.  Figure \ref{fig:lzs} presents modeled phase shifts for ground state potassium atoms with several different velocities and five different spin states. Figure \ref{fig:lzs} illustrates how spin-dependent dispersion compensation works, and how it can make $\LZL \neq \LZ$.

\begin{figure}[t]
\begin{center}
\includegraphics[width=9cm]{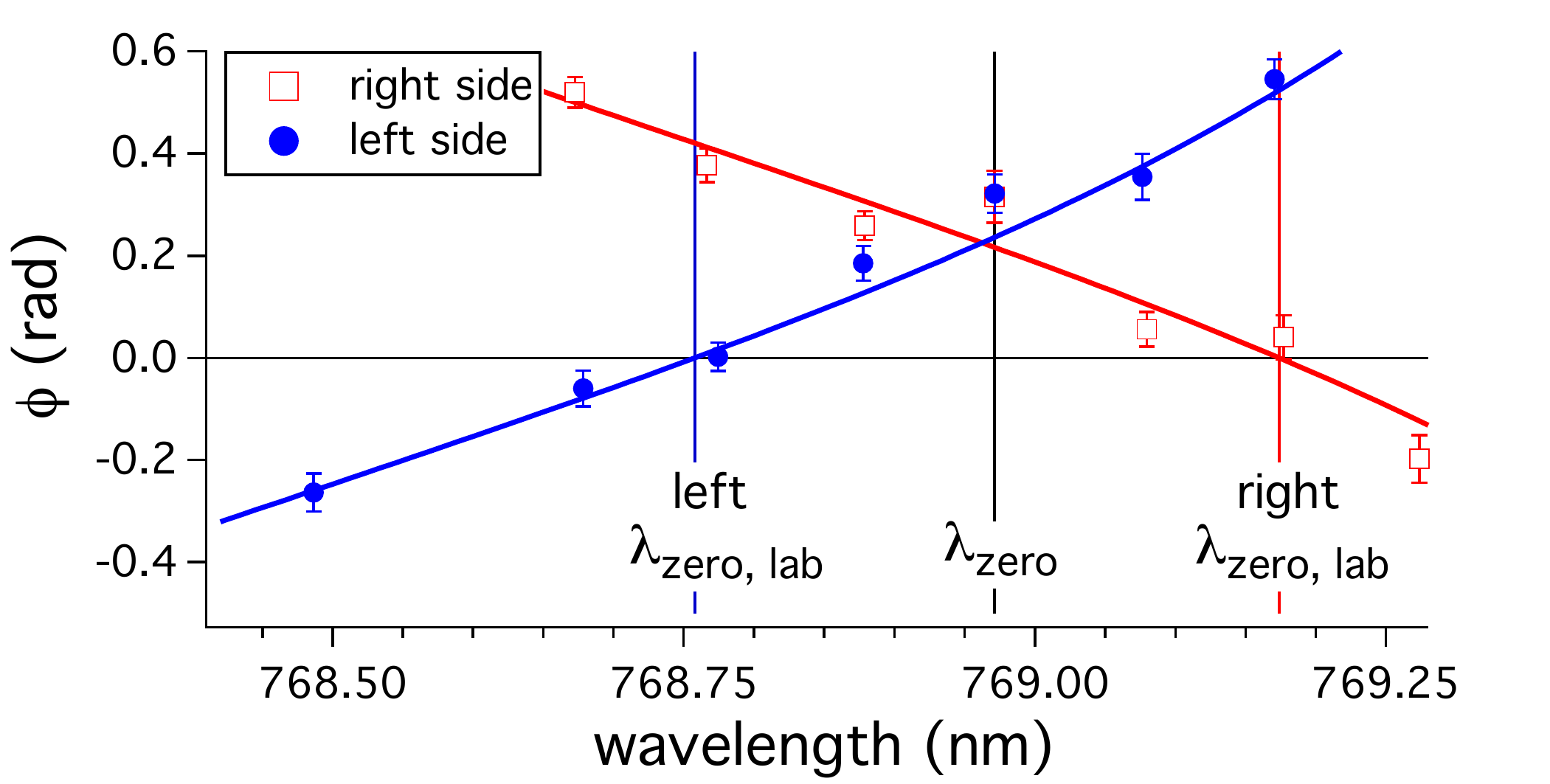}
\caption{Measured light-induced phase spectra $\phi\left(\lambda\right)$ using elliptically polarized light and a magnetic field parallel to the optical $k$-vector. The open square red data show $\LZL = 768.758(15)$ nm  when the laser beam is on the right side of the atom interferometer, and the solid circle blue data show $\LZL = 769.174(7)$ nm  when the laser beam is on the left side of the atom interferometer.  Each data point is the average of 40 five-second files and the error bars show the standard error of the mean. Broad band radiation from the tapered amplifier caused a systematic shift of 15(5) mrad that we accounted for in the $\phi$ data shown. The red and blue curves show the theory using Eqs. (\ref{eq:phase}) - (\ref{eq:light}) with an additional average over the width of the atom beam. For these data, the grating tilt $\theta$ was -20(5) mrad.}
\label{fig:lzshift}
\end{center}
\end{figure}

The way $\Phi_S$ affects the light-induced phase $\phi(\lambda)$ leads to several testable predictions that we experimentally verified. Equation (\ref{eq:cphase}) led us to predict a new wavelength $\LZL$ for which $\phi$ is zero.  A simulation of this prediction is shown in Fig.~\ref{fig:lzs}, and data demonstrating +203 to -213 pm shifts in $\LZL$ are shown in Fig.~\ref{fig:lzshift}.

Higher irradiance on the left interferometer path when looking from the source towards the detector would cause a longer $\LZL$ (if the grating tilt were zero, so that $\Phi_{a} = 0$).  This is because attraction towards light on the left compensates for $\Phi_S$ in the northern hemisphere, and only spin states with roots in $\alpha(\omega)$ at longer wavelengths are attracted to light at $\LZ$.  These states therefore contribute with more weight to $\phi\left(\LZ\right)$ because of dispersion compensation.  On the other side, if the irradiance is stronger on the right-hand interferometer path, then repulsion from the light compensates for $\Phi_S$, and spin states with roots in $\alpha(\omega)$ at shorter wavelengths contribute more to $\phi(\LZ)$. Grating tilt $\theta$ and the gravity phase $\Phi_{a}$ complicate this picture. In our experiment, the dispersion $d\Phi_{a}/dv$ is opposite and slightly larger in magnitude than the dispersion $d\Phi_{S}/dv$, so higher irradiance on the left path of the atom interferometer causes a shorter $\LZL$. Figure~\ref{fig:lzshift} shows data verifying this prediction.

\begin{figure}[b]
\begin{center}
\includegraphics[width=9cm]{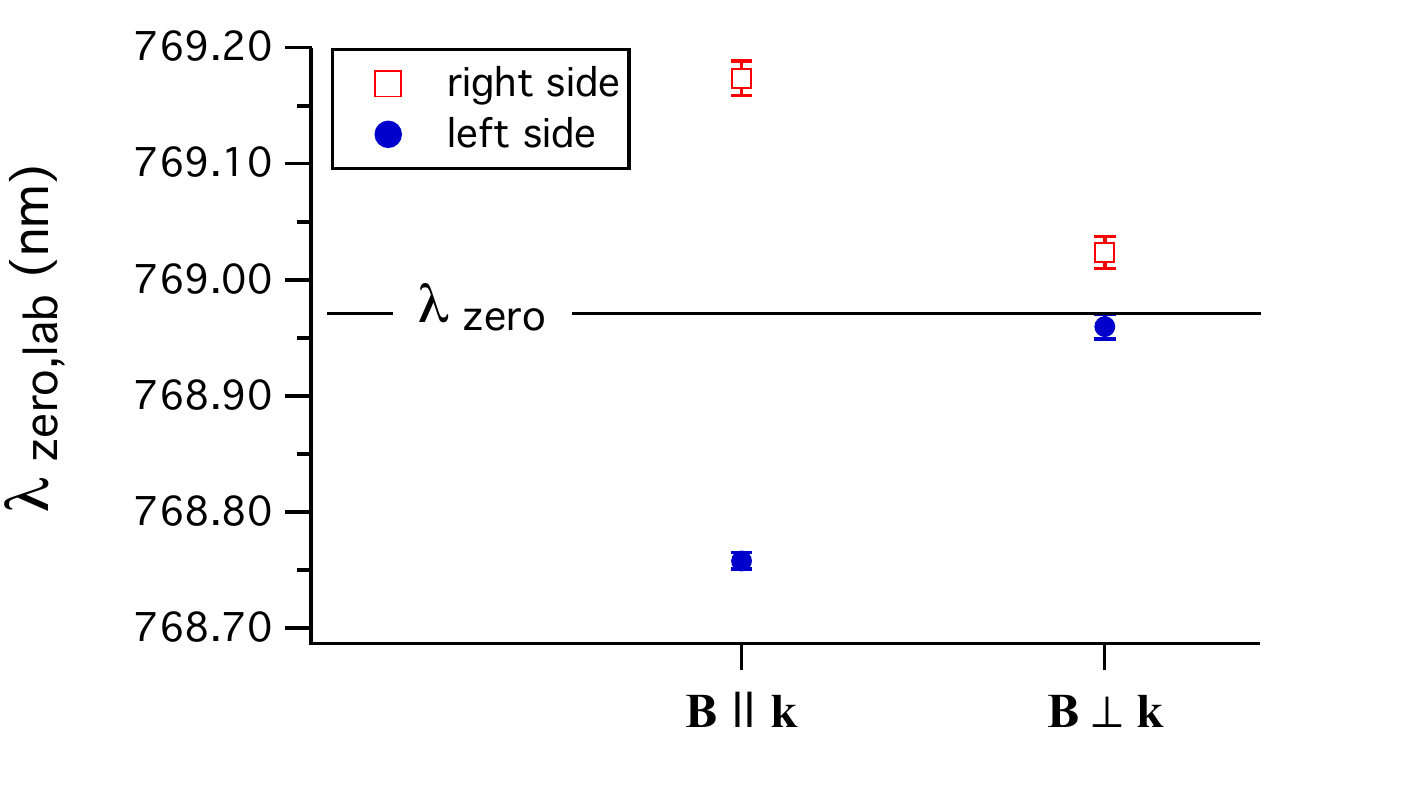}
\caption{Measured tune-out wavelengths for different orientations of magnetic field and irradiance gradients. Each data point comes from $\phi\left(\lambda\right)$ spectra such as those shown in Fig.~\ref{fig:lzshift}. For these data, the grating tilt $\theta$ was -20(5) mrad.}
\label{fig:blz}
\end{center}
\end{figure}

We predict that the wavelength difference $\Delta \lambda = \LZL - \LZ$ will not change if the optical $k$-vector reverses direction, nor if the optical circular polarization reverses handedness, nor if the magnetic field parallel to the optical $k$-vector reverses direction.  None of these reversals change the fact that a potential gradient that is attractive towards the left side (or repulsive from the right side) is needed to compensate for the Sagnac phase dispersion in the northern hemisphere.  Therefore, the magnitude $|\Delta \lambda|$ can increase if the laser is simply reflected over the atom beam path.  We tested this prediction by constructing an optical cavity with plane mirrors to recycle light so that the same interferometer path is exposed to upward and downward propagating laser beams for several passes.  This increased the magnitude of $\phi\left(\LZ\right)$ as predicted.

External magnetic fields also affect $\LZL$.  A uniform magnetic field parallel or anti-parallel to the optical $k$-vector maximizes the sensitivity to optical polarization.  Alternatively, a magnetic field perpendicular to the optical $k$-vector reduces $\Delta \lambda$ because the atomic spin states precess about the field so the resulting spin-dependent differences in light shift time-average to zero.  Data in Fig.~\ref{fig:blz} show that $\LZL$ is closer to $\LZ$ when we apply a perpendicular magnetic field. Residual differences between $\LZL$ and $\LZ$ are due to imperfect alignment of the magnetic field perpendicular to the $k$-vector and the limited (15 G) strength of the magnetic field.

Based on the work presented thus far, deducing $\Omega$ from measurements of $\Delta \lambda$ is challenging because it requires knowing the magnetic field, the laser power, laser polarization, laser beam waist, and the atom beam velocity spread.   To solve this problem, we used a static electric field gradient to induce additional phase shifts that mimic the effect of auxiliary rotation on the atom interferometer (to first order in $v$). A measurement of light-induced phase shift as a function of electric-field induced phase shift can serve to calibrate the relationship between $\Delta\lambda$ and $\Omega$. Furthermore, we can determine the absolute rotation rate of the laboratory by measuring the additional phase shift needed to make $\LZL = \LZ$. The phase due to a static electric field gradient is
\begin{equation}
\Phi_{\nabla E} = \frac{\alpha(0)}{2\hbar v}\int{s\cdot\frac{d}{dx}E^{2}dz}
\label{eq:gE}
\end{equation}
where $\alpha(0)$ is the static electric dipole polarizability \cite{Hol10}.  The observed phase shift for the ensemble of atoms due to an electric field gradient, $\phi_{\nabla E}$, is calculated using Eq.\nobreak~(\ref{eq:cphase}) with $\Phi_{\nabla E}$ added to $\Phi_{\textrm{total}}$ (and $\Phi_{L} = 0$). This phase shift can compensate for the dispersion in the Sagnac phase uniformly for all atomic spin states.

\begin{figure}[t]
\begin{center}
\includegraphics[width=9cm]{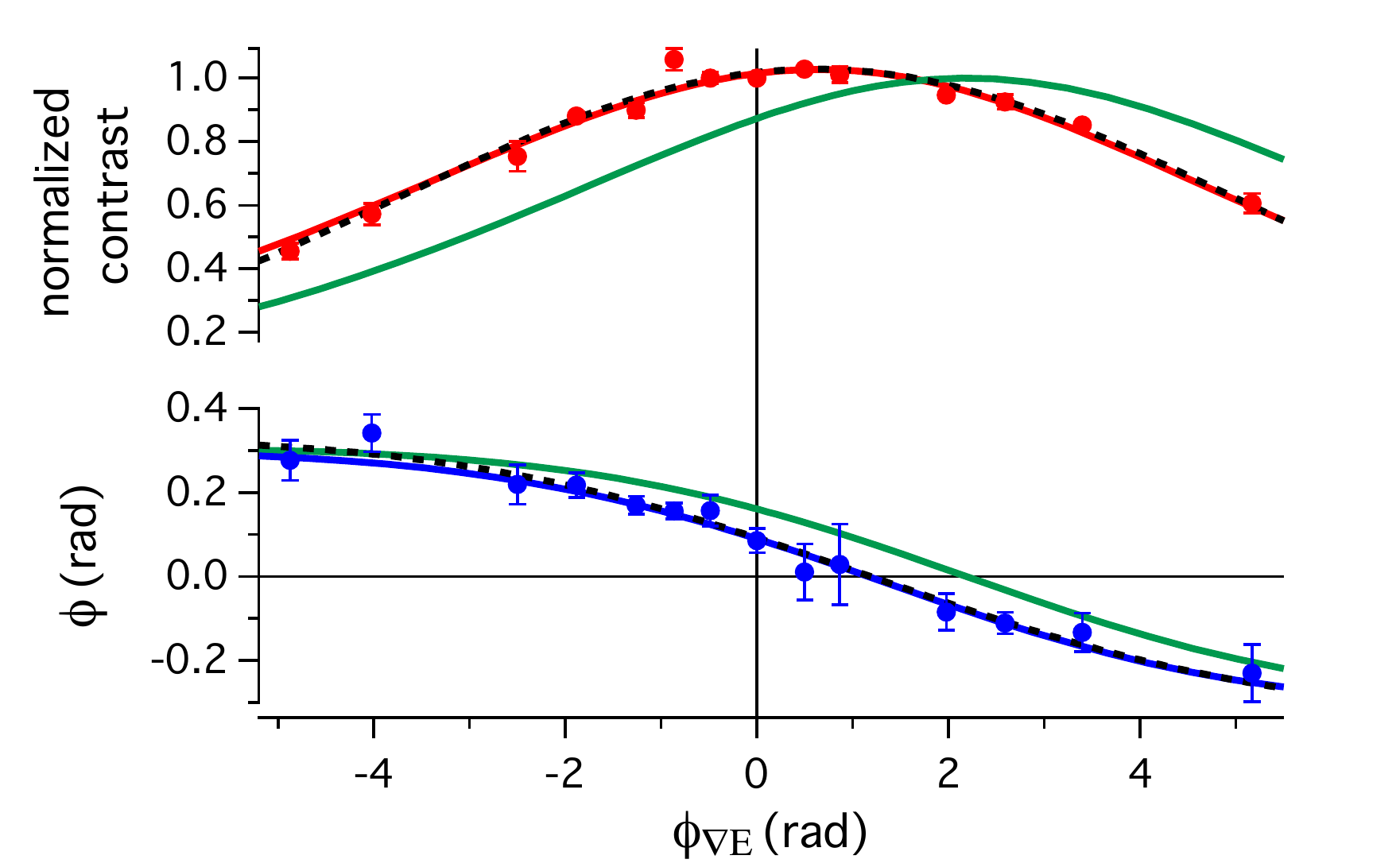}
\caption{(Top) Contrast data as a function of phase shift induced by an electric field gradient $\phi_{\nabla E}$. A Gaussian fit (dashed black) to the red data shows that a maximum in contrast occurs at $\phi_{\nabla E}= 0.6(2)$ rad due to dispersion compensation. The solid red curve shows the theory using Eqs. (\ref{eq:phase}) - (\ref{eq:gE}) with $\Omega = 0.6 \Omega_{e}$.  (Bottom) Light-induced phase shift $\phi$ as a function of $\phi_{\nabla E}$, using light at $\LZ = 768.971$ nm. An error function fit (dashed black) to the blue data shows the root $\phi_{\nabla E}^{\textrm{root}}$=1.2(3) rad. The solid blue curve shows the theory using Eqs. (\ref{eq:phase}) - (\ref{eq:gE}) with $\Omega = 0.4 \Omega_{e}$. For these data, the grating tilt $\theta$ was -10(2) mrad. The solid green curves show contrast and phase theory for $\Omega = 0$, but the same $\theta = -10$ mrad.}
\label{fig:lzge}
\end{center}
\end{figure}

In Fig.~\ref{fig:lzge}, we show that $\phi(\LZ)$ depends continuously on $\phi_{\nabla E}$, just as $\Delta\lambda$ would on $\Omega$.  Specifically, $\phi(\LZ)$ is the phase shift caused by light at $\LZ$. The data in Fig.~\ref{fig:lzge} are obtained by alternately turning $\nabla E$ on and off, blocking and unblocking the laser, and then repeating the process with a new $\nabla E$ strength.  Importantly, the root in phase $\phi(\LZ)$ at $\phi_{\nabla E}^{\textrm{root}}$ occurs when the electric field gradient compensates for dispersion in $\Phi_S$ and $\Phi_a$.  We can interpret this  condition mathematically as
\begin{equation}
\frac{d}{dv} \left( \Phi_S + \Phi_a + \Phi_{\nabla E} \right) = 0
\label{eq:ddv}
\end{equation}
and then it becomes unnecessary to know the laser power or to perform the integral over velocity shown in Eq.~(\ref{eq:cphase}) for reporting $\Omega$.  Using the approximation $  \Phi_{\nabla E} = \phi_{\nabla E}^{\textrm{root}} (v_0/v)^2  $ we find
\begin{equation}
\Omega =      - \frac{ d_g v_0 \phi_{\nabla E}^{\textrm{root}} }{2 \pi L^2} - \frac{ g \sin(\theta) }{v_0}.
\label{eq:omega}
\end{equation}
Equation~(\ref{eq:omega}) does not include $\Phi_{L}$ because when Eq.~(\ref{eq:ddv}) is satisfied there is no net dispersion to break the symmetry; so including $\Phi_{L}(\LZ)$ in Eq.~(\ref{eq:cphase}) produces zero ensemble phase shift $\phi$.  The fact that Eq.~(\ref{eq:omega}) does not include $\Phi_{L}$ is convenient because now we can use light at $\LZ$ to measure $\Omega$ without precise knowledge of the laser spot size, polarization or irradiance, or the resultant slope $d \phi / d \lambda$.  Those factors affect the precision with which we can find the root ($\phi_{\nabla E}^{\textrm{root}}$), but not the value of the root. We also emphasize that an electric field gradient can be used to increase the dynamic range of our gyroscope.   

To report $\Omega$, we measured $\phi_{\nabla E}^{\textrm{root}} = 1.2(3)$ rad with data in Fig.~\ref{fig:lzge}, we measured $v_0 = 1585 (10)$ m/s using phase choppers \cite{Hol11}, and we measured $\theta = -10(2)$ mrad by comparing the nanograting bars to a plumb line. We find $\Omega = 0.4(2) \Omega_e $, which can be compared to the expected value 0.5 $\Omega_e$ (the vertical projection of $\Omega_e$ at our latitude of 32$^{\circ}$ N).  In Fig.~\ref{fig:lzge}, we also show how $C_{\textrm{off}}$ depends on $\phi_{\nabla E}$. The phase $\phi_{\nabla E}^{\textrm{max}C}$ that maximizes contrast is another way to find the static electric field gradient that compensates for dispersion in the Sagnac phase and acceleration phase.   The value of $\phi_{\nabla E}^{\textrm{max}C}$ = 0.6(2) rad leads to $\Omega = 0.6(2) \Omega_e $. The dominant source of error in our experiment was the measurement of the nanograting tilt. Discrepancy between $\phi_{\nabla E}^{\textrm{max}C}$ and $\phi_{\nabla E}^{\textrm{root}}$ indicates a systematic error, possibly caused by de Broglie wave phase front curvature induced by the laser beam \cite{Hro14}, optical pumping, magnetic field gradients, or the broad band component of our laser  spectrum.

The shot noise limited sensitivity of our atom interferometer gyroscope can be estimated from the fact that $\phi(\LZ)$ changes by 0.22 radians due to 0.53 $\Omega_e$, and the statistical phase noise is $\delta\phi = (2/N)^{1/2}[(1/C_{\text{off}})^{2}+(1/C_{\text{on}})^{2}]^{1/2}$ is 0.06 radians/$\sqrt{\textrm{Hz}}$ for our experimental values of $C_{\text{off}} = 0.2$, $C_{\text{on}} = 0.08$ and $N$ = (100,000 counts/sec)$\times t$.  This indicates a sensitivity of $0.2 \Omega_e /\sqrt{\textrm{Hz}}$ for measurements of rotation with respect to an inertial reference frame, which is competitive with methods presented in  \cite{Gus97, Dic13, Dur06, Can06, Lan12}. 

To make a more sensitive gyroscope the scale factor  $\phi(\LZ)/\Omega$ can be somewhat increased by using more laser power and a broader velocity distribution.  However, a limit to the sensitivity arises from balancing the benefit of an increased scale factor against the cost of increased statistical phase noise.  This compromise occurs because maximizing the scale factor, $\phi(\LZ)/\Omega$, requires significant contrast loss from the two mechanisms described by Eq.~(\ref{eq:cphase}): first, averaging over the spread in $\Phi_S$ (which is affected by $\sigma_v$) and second, averaging over the distribution in $\Phi_L$ (which is affected by the laser power and polarization).  Optimizing $\sigma_v$ and laser power can increase the sensitivity (for the same flux and contrast) to $0.05 \Omega_e /\sqrt{\textrm{Hz}}$ for $\Omega$ measurements.

This work also indicates how to make measurements of $\LZ$ more independent of $\Omega$. Experiments are less sensitive to $\Omega$ if they use linearly polarized light, a narrow velocity distribution, a perpendicular magnetic field, and an additional dispersive phase such as $\Phi_{\nabla E}$ to compensate for $\Phi_S$.  For example, the $\LZ$ measurements in reference \cite{Hol12} were not significantly affected by $\Omega$ because there was minimal contrast loss at $\LZ$.  Specifically, the sharp velocity distribution ($v_0/\sigma_v = 18$) made dispersion in $\Phi_S + \Phi_a$ reduce $C_0$ by less than 1\%, and $\Phi_L$ only reduced $C$ by 4\% of $C_{0}$ so shifts in $\LZL$ were less than 1 pm in \cite{Hol12}.   To increase sensitivity to $\Omega$ for measurements reported here, in Figs.~\ref{fig:lzshift} - \ref{fig:lzge}, we used a broad velocity distribution ($v_0/\sigma_v = 7$) so $\Phi_S + \Phi_a$ reduced $C_0$ by 8\%, and we also used a large irradiance gradient with circular polarization that reduced $C$ by 40\% of $C_0$.

In summary, an atom beam interferometer with multiple atomic spin states enabled us to demonstrate systematic shifts in tune-out wavelength measurements ($\LZL$) that are larger than 200 pm due to rotation and  acceleration.  Then, we used the phase induced by light at a theoretical tune-out wavelength $\phi(\LZ)$ as a function of an additional dispersive phase $\phi_{\nabla E}$ applied to report the rotation rate of the laboratory with an uncertainty of 0.2 $\Omega_e$. This work is a new application for tune-out wavelengths, paves the way for improving precision measurements of tune-out wavelengths, and demonstrates a new technique for atom interferometer gyroscopes.  The spin-multiplexing techniques demonstrated here may find uses in other atom \cite{Pet13,Lom14} and neutron \cite{Has11,Haa14} interferometry experiments, and also in NMR gyroscopes and NMR spectroscopy.  

This work is supported by NSF Grant No. 1306308 and a NIST PMG. R.T. and M.D.G. also thank NSF GRFP Grant No. DGE-1143953 for support. We also thank Professor Brian P. Anderson for helpful discussions.

\bibliography{refs}
\end{document}